# Understanding the Trajectories of Population Decline Across Rural and Urban Europe: A Sequence Analysis

Niall Newsham[1,*], Francisco Rowe[1]

[1]*Geographic Data Science Lab, University of Liverpool, Liverpool, UK*
*Correspondence: n.newsham@liverpool.ac.uk*

## Abstract

Population decline is projected to become widespread in Europe, with the continental population set to reverse its longstanding trajectory of growth within the next five years. This represents unfamiliar demographic territory. Despite this, literature on decline remains sparse and our understanding porous. Particular epistemological deficiencies stem from a lack of both cross-national and temporal analyses of population decline. This study seeks to address these gaps through the novel application of sequence and cluster analysis techniques to examine variations in population decline trajectories since 2000 in 696 sub-national areas across 33 European territories. The methodology allows for a holistic understanding of decline trajectories capturing differences in the ordering, timing, magnitude and spatial structure of population decline. We identify a typology of population decline distinguishing seven distinct pathways to depopulation and chart their geographies. Results revealed differentiated pathways of depopulation in continental sub-regions, with consistent and rapid declines in the east, persistent but moderate declines in central Europe, accelerating declines in the south and decelerating population declines in the west. Results also revealed differentiated patterns of depopulation across the rural-urban continuum, with urban and populous areas experiencing deceleration in population decline, while population decline accelerates or stabilises in rural areas. Small and mid-sized areas displayed heterogeneous depopulation trajectories, highlighting the importance of local contextual factors in influencing trajectories of population decline.

## Introduction

Europe is projected to become the first continent to undergo a unique demographic transition - population decline, or depopulation[1]. Most recent estimates from the United Nations predict the continental population to be in a state of decline by 2025 (UN 2019). This trajectory of depopulation is set to persist for the remainder of the 21st century, defining European demography. By 2100, Europe's population is forecast to shrink by around 120 million inhabitants, or by 15% (UN 2019 - medium variant). Population decline is not, however, expected to occur uniformly across the continent as significant differences in the rate and direction of population change are expected to continue, further exacerbating regional and country demographic imbalances (Eurostat 2020, ONS 2020, UN 2019). At present, population decline is under way in 17 of 48 European countries, the majority of which confined to Eastern and Southern Europe. In the near future, decline is expected to spread to all regions of Europe with 33 countries set to undergo decline by 2050 (UN 2019). Depopulation at this scale is a previously unrecognised demographic phenomenon and will thus impose a wealth of novel challenges (Coleman & Rowthorn 2011, Clements et al., 2015, Franklin 2020).

Contemporary population decline is often portrayed in such a way that emphasises contrasting trajectories of population change between rural and urban areas (Coleman & Rowthorn 2011, Johnson & Lichter 2019, Franklin 2020). Though it is true that decline is more prevalent in rural areas (EPSON 2017), recent empirical evidence demonstrating population growth in certain rural areas (Pužulis & Kūle 2016, Salvati 2018, Wojewódzka-Wiewiórska 2019) and decline in urban areas (Turok & Mykhnenko 2007, Wolf & Weichmann 2018) would suggest that this notion is over-simplified. Rather, depopulation processes are sensitive to local contextual specificities (Haase et al., 2016, Doignon et al., 2016) and extend to all areas along the rural-urban continuum

Though inherently a demographic process, population decline remains under researched. As a result, we lack a sufficient understanding of the spatial and temporal dynamics of the phenomenon. The majority of research focuses on small scale case studies and provide only a highly specific account of localised population declines (Viñas 2019, Spórna et al., 2016, Szmytkie 2016). This issue is compounded by an intense rural-urban dichotomy that persists within relevant research, with studies exclusively concerned with either rural (Kuczabski & Michalski 2013, Wojewódzka-Wiewiórska 2019, Pužulis & Kūle 2016) or urban decline (Haase et al., 2016, Wolff & Weichmann 2018). Rendering a holistic understanding of European depopulation is needed but currently difficult to attain. Additionally, the temporal dynamics of decline are seldom considered in research. Too often studies focus only on the outcome of depopulation and fail to acknowledge the process in which decline unfolds. As emphasised by Franklin (2020), population decline is both a process and an outcome. Also recognising this, Haase et al., (2016), call for a move towards process-oriented research into population decline. Population change is a path dependence process. Understanding past patterns of population change is key to build our knowledge about present and future population patterns (Patias et al. 2021) To our knowledge, no study has empirically analysed population decline processes in a comparative cross-national spatial framework across the rural-urban continuum. We aim to address this by identifying distinctive trajectories of sub-national population decline; establish differences in the sequence, timing and degree in these trajectories; and determine the spatially differentiated extent of depopulation across Europe. To these ends, we apply a novel methodology in sequence analysis to a dataset of annual population change capturing trajectories of depopulation in 696 sub-national areas covering the period 2000-2018. Created for the study of DNA sequencing, sequence analysis has been applied in population studies to analysis individual-level trajectories over the life course (e.g. Rowe et al. 2017; Backman et al., 2021). But it has rarely been used to examine the temporal evolution of spatial population trajectories, except for applications to understanding changes in the socio-economic composition of neighbourhoods (Patias et al., 2019; 2021). Sequence analysis has the potential to expand our understanding of population decline by embedding individual population changes within a wider framework of population trajectories; that is, a sequence of interlinked changes experienced by an area.

---

[1]*Population decline and depopulation are used interchangeably throughout this article.*

Next, we first review existing literature on sub-national population decline in Europe and introduce the urban differentiation model (Geyer & Kontuly 1993) as a useful framework to contextualise and analyse trajectories of population decline across the rural-urban continuum. We then describe in detail the methodologic procedure, explaining the application of sequence analysis. Our results are then presented, followed by a discussion and a concluding section.

## Literature Review

Europe is facing unprecedented population declines. Across the continent, in all countries, fertility rates no longer sustain population growth. Instead fertility rates currently lie below the replacement level of 2.1 births per woman (UN 2019) and promote natural population declines. Though not yet occurrent in all countries of Europe on a national scale, population decline is particularly pronounced in Eastern and Southern European countries. With population momentum waning in growing countries, population decline is averted through positive net migration streams (Sobotka & Fürnkranz-Prskawetz 2020). Migration, however, intrinsically promotes depopulation in the places of origin (Franklin 2020) and can accelerate population declines. Though, it remains unclear how the effects of COVID-19 will shape the demographic future of Europe, early evidence of crude birth rate reductions (Aassve et al., 2021) combined with disrupted migration flows points to an immediate future with less people than present.

### Consequences of Decline

Population decline is expected to bring about unprecedented challenges across society. Inherently a local issue, population decline poses considerable threats to communities.

Particularly, areas in decline are faced with shrinking local tax revenues, impacting the provision of vital services (Carbonaro et al., 2018) including transport (Franklin et al., 2018). Areas in decline are also less attractive to prospective businesses and face a stagnation in regards to opportunity. Additionally, as a demographic process, population decline will alter the composition of an area. With this, changes to age structure, diversity, and attitudes will be likely observed (Franklin 2020). Such changes will bear social and economic implications of their own, namely income inequalities (Bellman et al. 2018) and the reduction in area competitiveness (Poot 2008) potentially fuelling further population losses. In short, decline will have a profound impact on the way local areas are experienced and perceived. On a national scale, depopulation poses challenges that are closely associated with aging populations. Particularly, increases in expenditure on pensions and health services (European Union's Economic Policy Committee 2010) could accompany increasing public debts, severe cuts in other spending, and tax increases (Clements et al., 2018). Population aging, and decline, threaten future economic productivity and growth due to labour and skills shortages (see Bloom et al., 2010, Clements et al., 2015).

**Sub-national Decline**

The risk of population decline is not equal across sub-national regions. General patterns in the geographic distribution of depopulation show that rural areas are currently dominating the landscape of European population decline (Eurostat 2020) with a growing number of rural regions experiencing depopulation in recent years (Dax & Fisher 2018). Rural population decline is more commonplace in central and eastern Europe than in western Europe (EPSON 2017), although it extends to all regions of the continent (Eurostat 2020). It should be noted, however, that not all rural areas are prone to population decline. Rather there exists fragmented country specific evidence showing growth within rural areas situated in close proximity to large urban areas. This has been demonstrated in Lativa (Pužulis & Kūle 2016), Lithuania (Ubarevilien et al., 2016) and Poland (Wojewódzka-Wiewiórska 2019). Rural decline is perhaps then more pronounced in remote and peripheral rural areas, though it remains unclear if this is a continental-wide pattern. In contrast, urban areas are continuing to record population growth (Eurostat 2020) and are less likely to experience depopulation. Urban areas are not, however, exempt from depopulation. Rather, almost half of all European cities had experienced a period of decline between 1990-2010 (Wolf & Weichmann, 2018). Certain cities, however, appear more likely to experience population decline than others. Geographically, incidences of urban population decline are three times more prevalent in the Eastern EU-13 nations than in the Western 15 (European Commission 2016). Recent research also suggests that position within the urban hierarchy is important in determining population decline outcomes. In an empirical test of the association between city size and population decline, Kabisch & Haase (2011) find that smaller sized city agglomerations were more likely to experience population decline than large and mid-sized agglomerations. The propensity for population growth in European capital cities reinforces this notion (Lutz et al., 2019, Karachurina & Mkrtchyan 2015). Despite this, anomalies to this trend are evident, for example recent population declines observed in Athens and Thessaloniki, the largest urban agglomerations in Greece (Salvati 2018).

**Trajectories of Population Decline**

Population decline is both an outcome and processes (Franklin 2020). Research concerning European depopulation too often only considers the measured outcome, i.e. reduction in numbers of population, and ignores the process of decline over time (Turok & Mykhnenko 2007). This leaves many crucial questions unanswered; *how have areas experienced population decline over time? Are population declines sudden and rapid, or are they longstanding and gradual? Do trajectories of population decline differ between countries or between rural and urban regions? are decline trajectories permanent or can they be reversed? What may influence trajectories of population decline?* Understanding the temporal evolution, pace and extent of decline is vital to developing appropriate policy responses, improving population projections and advancing demographic theory concerning decline.

From existing research, we understand the existence of divergent trajectories of population decline both within and between countries (Haase et al., 2016, Wolff & Weichmann 2018). However, such studies consider only urban regions and limit our understanding of continental wide depopulation processes. A comprehensive, cross-national, study into sub-national depopulation processes is desirable for numerous reasons. Firstly, the extent of population decline in individual country and sub-national regions can be more effectively assessed when considered within a cross-national context. Secondly, cross-national comparative analyses have the potential to demonstrate new insights into population decline through uncovering nuanced patterns and processes. Finally, cross-national comparisons provide a more rigorous test-bed for theories concerning depopulation. We consider developments in the rate of population change as a measure of depopulation processes and study these in this paper.

**Urban Differentiation**

In absence of a comprehensive understanding of temporal depopulation processes, the urban differentiation model (Geyer & Kontuly, 1993) provides a useful theoretical framework in which trajectories of population decline can be contextualised. Yet, we recognise that this model was not devised to study trajectories of population decline. Particularly the model assumes that national human settlement systems will follow a predefined sequence of cascading population changes driven mainly by net migration outcomes. The urban differentiation model postulates that groups of large, intermediate-sized and small cities go through successive periods of fast and slow population growth or decline, resulting in stages of population concentration or deconcentration (Geyer 1996). In a first stage, fast population growth is assumed to occur in primate cities, with smaller cities and rural settlements losing population, resulting in urbanisation or population concentration. This is followed by a second phase in which population growth is still concentrated in primate cities but occurs at a more modest pace, and intermediate cities start experiencing moderate population growth. In a third stage, population growth mostly occurs in intermediate cities and small cities start recording population rise, while the rate of population change decreases in main urban centres. During a fourth phase, population deconcentration in the form of counter-urbanisation becomes the dominant pattern reflecting concentration of population growth in small cities, population decline in major cities and decreasing population growth in intermediate urban areas**.** The model assumes that the process restarts from the first stage with a new phase of population concentration occurring as population growth concentrate in primate cities. Though, population changes during the subsequent cycles are expected to be less intense in magnitude than those during the initial phase.

Given that population dynamics are spatially varying, the urban differentiation model offers a useful framework to analyse changes in population over time. Particularly, the model recognises fluctuations in rates of population change and the divergence of trajectories across the urban-rural continuum. Should the model hold true, we would expect to find a structured set of population decline trajectories representing the individual stages of population concentration and deconcentration outlined in the model. For example, if a country is experiencing concentrated population growth in chief urban areas, we should observe trajectories of accelerating depopulation in rural and small cities. Similarly, if a country is experiencing urban population deconcentration or population decline, annual growth should be observed in rural or small cities. However, we postulate that trajectories of population decline are more complex and mediated by local contingencies so we do not expect that national settlement systems to follow these predetermined sets of transitions.

## Methodology

### Data

We collected annual population data involving 2,035 areas in 43 European territories over an 18-year period between 2000 and 2018. We used NUTS (Nomenclature of Territorial Units for Statistics) 3 units corresponding to the smallest regional unit in the Eurostat territorial classification system. A database was assembled using data from Eurostat and national statistics offices. Eurostat data were used as the base and supplemented with national statistics data for European territories excluded from Eurostat, or where data were incomplete. Appendix 1 provides a breakdown of data sources by territory. Data constraints do not allow to expand the period of analysis and build a longer time series. Given these constraints, we acknowledge that our data may not capture the start of the contemporary trajectory of population decline in some areas (e.g. Martí-Henneberg 2005).

From our database, we identified a total of 736 sub-national areas that experienced an overall population decline from 2000 to 2018. These areas are characterised by their relative population losses and are the focus of this study. A significant challenge of building a spatial panel dataset is presented as geographic boundaries change over time (Rowe 2017). We were largely able to mitigate the impact of this through harmonising these changes based on Eurostat's correspondence files which provides a historical record of these changes. Yet, 40 areas were missing data for a significant portion of the time series and were removed from the analysis. These areas are exclusively located within Germany (25) and Poland (15) (see Appendix 1). Therefore, our analysis concerns the trajectories of population decline in a total of 696 sub-national areas in 33 European territories. Of these areas, 131 have an incomplete time series on the basis of missing data for a single year (i.e. 2000 or 2001). We decided to include these areas in our analysis. Our results are robust to their inclusion. Additionally, we used data on the population distribution by settlement type to construct a rural-urban typology from NUTS-3 areas from Eurostat, and supplement with national statistics data for areas not included in the Eurostat dataset (Appendix 1).

**Method**

We developed a four-stage methodological strategy to define and analyse pathways of population decline, as shown in Figure 1. We first computed annual region-specific rates of population change and analysed their distribution across Europe to identify a set of thresholds and used them to classify regions. Second, we applied a sequence analysis technique, known as optimal matching, to measure differences in the temporal profile of population change across individual regions. Third, we used these measures to define a typology of population decline trajectories using a hierarchical clustering procedure. In a fourth stage, we examined the geographic distribution of these trajectories and differences across the urban-rural continuum. Next, we describe the implementation of each of these stages.

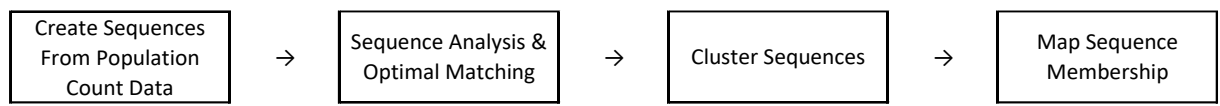

Figure 1 – Methodological procedure

*Stage 1*

To enable the implementation of sequence analysis, we classified population count data into distinct categorical states or classes as sequence analysis requires longitudinal categorical data as an input. We computed the annual rate of percentage population change for individual areas and used these rates to differentiate areas of high, moderate and stable population growth and decline. The annual rate of population change was calculated as follows:

$$\text{Rate of Population Change} = \frac{P(t2) - P(t1)}{P(t1)} \quad (1)$$

Where P(t1) is the population at year t and P(t2) is the population at t+1.

We distinguished five classes of population change, capturing various magnitudes of population growth, decline and stability (Table 1), and representing the relative rate of annual population change. The trajectories of population change are represented by the chronological ordering of these classes. To determine appropriate thresholds for the definition of classes presented in Table 1, we analysed the distribution of the annual rate of population change, particularly its central tendency and dispersion over the 2017-18 period. The median (0.303) was used as middle point and threshold to define the boundary between patterns of population *Stability* and *Moderate Growth*. The standard deviation (0.990) was used as upper boundary to define trends of *Moderate Growth*. The resulting categorisation closely aligns with existing definitions of population decline (UN 2019, Wolff & Weichman 2018). The additive inverse of these thresholds were used to define patterns of *Stability, Moderate Decline* and *Decline*. We applied the resulting classification to all area-year observations in our data set to identify how the pattern of population change evolves for individual areas.

| State | Definition (annual % change) |
|---|---|
| Decline | ≤ -0.99 |
| Moderate Decline | > -0.99 & ≤ -0.3 |
| Stability | > -0.3 & < 0.3 |
| Moderate Growth | ≥ 0.3 & < 0.99 |
| Growth | ≥ 0.99 |

Table 1 – Defined states of population change for sequence analysis.

*Stage 2*

Next, we measured the dissimilarity between individual sequences of population change, to identify similar types of trajectories in Stage 3. To this end, we used optimal matching (OM) which computes distances between sequences as a function of the number of transformations required to make sequences identical. Three transformation operations are used: insertion/deletion (indel) or substitution. Each of these operations have a cost, and the distance between two sequences is defined as the minimum cost to transform one sequence to another (Abbott & Tsay 2000). The greater the cost to make two sequences identical, the greater the dissimilarity and *vice versa*. The costs of indel and substitution operations are not equally weighted. By default, indel costs are set to 1 and substitution costs are empirically derived from transition rates between states. The cost of substitution is inversely related to the frequency of observed transitions within the data. This means that infrequent transitions between states have a higher substitution cost. For example, transitions from the state of decline to stability are rarer than decline to moderate decline, and so this is represented by a higher cost (Table 2).

A total of 129 areas in our data set have no observations for the year 2000, and 40 are also missing data for 2001. We considered missing observations as an additional class in our analysis and assigned a substitution cost of 2, the highest cost, to minimise the impact on our results. The intuition of this is: substitution for missing data is expensive so dissimilarity measures are rarely based on missing data, and if this happened, such observations are grouped into a single category (see below). Our weighting scheme, that is the cost of indel operations being less than substitutions, enables us to uncover differences in the sequencing of population changes according to our categories, rather than their timing. This is because indel operations produce a time shift between compared sequences in order to identify identical subsequences (Lesnard 2014). Indel operations favour the identical ordering of states irrespective of their position in the sequence (Lesnard 2010).

|  | Decline → | Moderate Decline → | Stability → | Moderate Growth → | Growth → | Missing → |
|---|---|---|---|---|---|---|
| Decline → | 0 | 1.684 | 1.897 | 1.949 | 1.852 | 2 |
| Moderate Decline → | 1.684 | 0 | 1.623 | 1.867 | 1.721 | 2 |
| Stability → | 1.897 | 1.623 | 0 | 1.499 | 1.718 | 2 |
| Moderate Growth → | 1.949 | 1.867 | 1.499 | 0 | 1.746 | 2 |
| Growth → | 1.852 | 1.721 | 1.718 | 1.746 | 0 | 2 |
| Missing → | 2 | 2 | 2 | 2 | 2 | 0 |

Table 2 – Substitution cost matrix depicting the costs assigned to each transition.

Stage 3

The resulting distance matrix from Stage 2 was used as an input for a cluster analysis, to produce a typology of population decline trajectories. The Ward's hierarchical ascending clustering algorithm (Ward 1963) was used. To determine the optimal number of clusters, a range of cluster quality measures were empirically evaluated using the WeightedCluster package in R (Studer 2013). These include Average Silhouette Width (ASW) which measures distances between clusters and within group homogeneity (Kaufman & Rousseeuw 1990); Hubert's Somers' D (HGSD) which is a measure of the clusters capacity to reproduce the distance matrix (Hubert and Arabie 1985); and, Point Biserial Correlation (PBC) which is similar to HGSD, but rather measures the capacity to reproduce the exact value of the distance matrix (see Hennig and Liao 2010). A graphical comparison of these metrics for a range of cluster solutions is provided (Appendix 2). From this, 7 clusters were decided as the optimal solution.

Stage 4

In a final stage, we analysed the spatial distribution of the population trajectories. We measured the geographic spread and concentration of these trajectories by country in Europe and examined differences in their incidence across the urban-rural continuum. For the latter, we analysed the distribution of population trajectories across areas by population size and settlement type. We classified areas by population size and share of rural population based on four population size categories - derived from the OECD/EC urban centre size classification (Dijkstra & Poelman 2012) - and three settlement types - determined by the proportion of an area's rural population, in accordance with the Eurostat Urban-Rural typology (Eurostat 2013) – see Table 3. Cross-tabulating these categories produces a rural-urban typology that accounts for different population sizes, enabling a more rigorous analysis of population decline processes by considering both area classification and size.

|  | XL | Large | Mid-Sized | Small |
|---|---|---|---|---|
| Population Size | > 1,000,000 | > 500,000 & < 1,000,000 | > 250,000 & < 500,000 | < 250,000 |

|  | Urban | Intermediate | Rural |
|---|---|---|---|
| Share of Rural Population | < 20% | >20% & <50% | > 50% |

Table 3 – Urban-rural typology

## Results and Discussion

**Overall Patterns of Decline**

Population decline has taken place in Europe since 2000. Yet, less is known about the magnitude and spatial distribution of these changes. Figure 2 reveals the extent to which population declines is unequally distributed across the continent. We observe a critical disparity in the extent of such declines between the east and west of Europe. Generally, the extent of decline is significantly more prevalent and severe within countries located east of Germany. Particularly, the greatest declines are observed within the Baltic and Balkan states where examples of sub-national areas exceeding 20% population decline are abundant.

Figure 2 also reveals the geographical spread and concentration of population decline within individual countries. Particularly, growth and decline contrasts are evident between northern and southern Italy (see Reynaud et al., 2020), western and eastern Germany, northern and southern Austria, eastern and western Portugal, northern-western and rest of Spain. Growth is typically observed in major urban areas, chiefly capital cities and its surrounding areas. In countries where depopulation is a prevailing feature, population growth is exclusively concentrated in these urban centres. This is true for Romania (Bucharest), Bulgaria (Sofia), Croatia (Zagreb), Latvia (Riga), Lithuania (Vilnius), Greece (Athens), Portugal (Lisbon and Porto) and Hungary (Budapest).

While our focus is on population decline, growth is acknowledged as the continuing dominant direction of population change in Europe, particularly within countries in the north and west of the continent. Of the 43 European territories investigated within this study, 5 have no sub-national areas experiencing depopulation: Belgium, Ireland, Malta, Norway, and Switzerland. Another 5 have not recorded a population decline though they comprise small nations with no sub-national regions recognised at the NUTS 3 level: Cyprus, Liechtenstein, Luxembourg, Montenegro and San Marino.

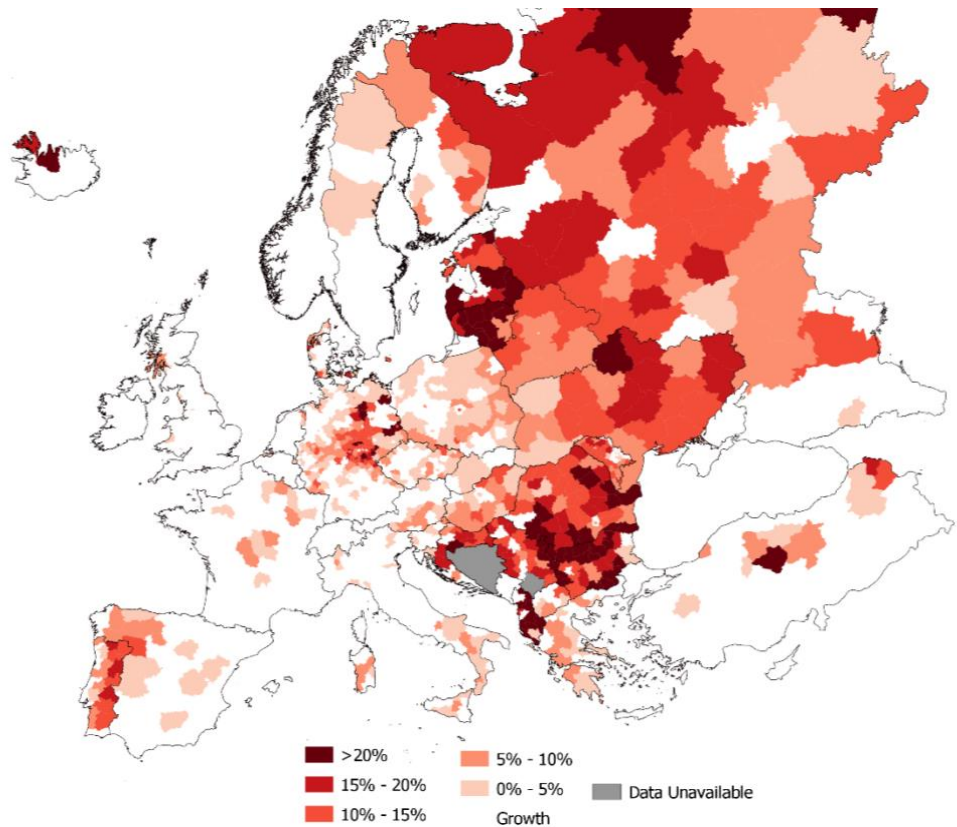

Figure 2 – Extent of population declines since 2000

**Trajectories of Population Decline**

These differences in the extent of sub-national population declines are underpinned by systematic differences in the pace and timing of depopulation. Figure 3 presents seven distinct pathways of depopulation that we identified through the application of sequence and clustering analyses as described in Stages 2 and 3 of our methodology. These pathways represent the seven distinctive ways in which population decline has unfolded across the continent.

Figure 3 presents three sets of plots: (1) state index plots, (2) state distribution plots, and (3) mean time plots for our seven trajectories of depopulation. Index plots display individual sequences, with each line representing an area and each colour denoting a class of population change from growth to decline. Reading horizontally, each line shows transitions between classes of population change over time, representing fluctuations in the rate and direction of population change. State distribution plots read vertically, showing the distribution of each class of population change for each year. Mean time plots indicate the average number of years that areas spend on each class. Table 4 complements these plots offering key summary statistics to describe each trajectory.

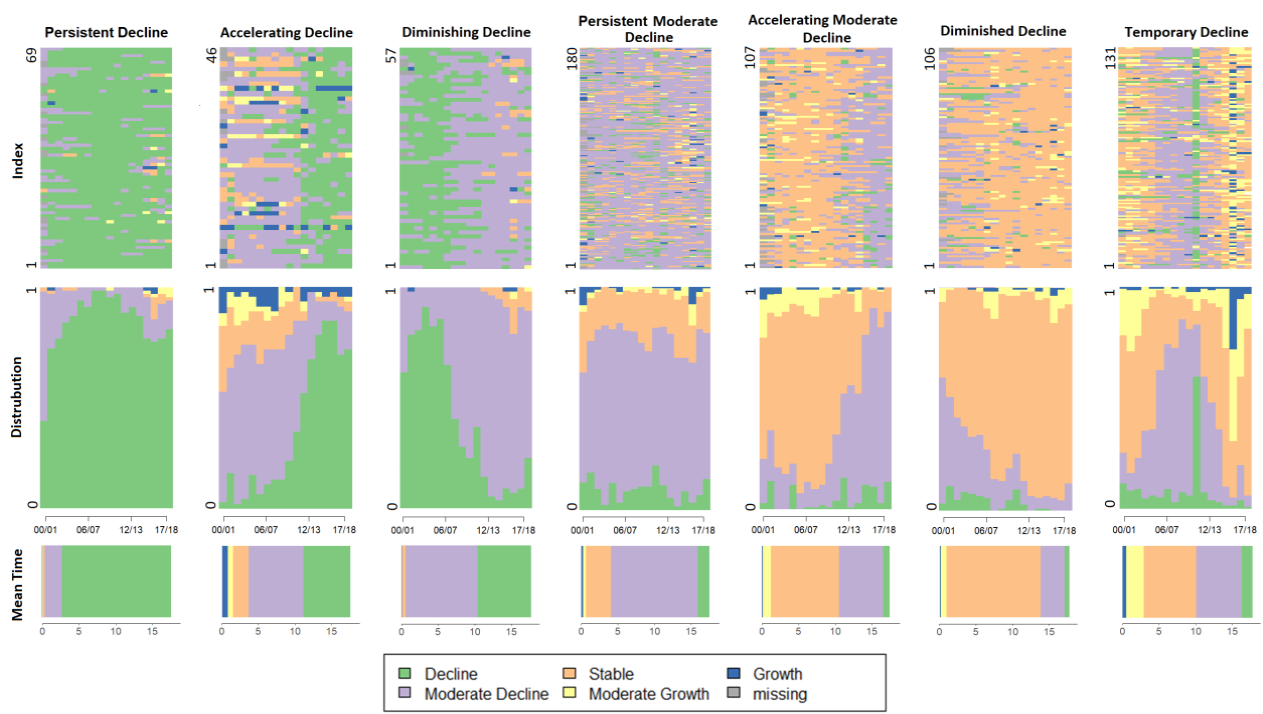

Figure 3 – Typology of European population decline

| | Cluster | | | | | | |
|---|---|---|---|---|---|---|---|
| | Persistent Decline | Accelerating Decline | Diminishing Decline | Persistent Moderate Decline | Accelerating Moderate Decline | Diminished Decline | Temporary Decline |
| n | 69 | 46 | 57 | 180 | 107 | 106 | 131 |
| n (%) | 9.91 | 6.61 | 8.19 | 25.86 | 15.37 | 15.23 | 18.82 |
| Population (2000) | 16,204,176 | 10,533,267 | 52,011,267 | 84,589,205 | 33,390,791 | 76,373,660 | 24,265,912 |
| Population (2018) | 12,356,331 | 8,957,866 | 44,190,272 | 76,724,267 | 31,440,886 | 73,562,720 | 23,289,173 |
| Decline | 3,847,845 | 1,575,401 | 7,820,995 | 7,864,938 | 1,949,905 | 2,810,940 | 976,739 |
| Decline (%) | 23.75 | 14.96 | 15.04 | 9.30 | 5.84 | 3.68 | 4.03 |

Table 4 – Measurement of the extent of decline within each population decline trajectory

*Persistent Decline (9.91%)*: This trajectory is composed of areas displaying an unwavering pattern of population decline of the highest magnitude (≤-0.99% per-annum, see Table 1). Areas tend to follow a consistent pattern of population decline. Very few transitions between classes are observed. Relative population loss in areas within this trajectory is the greatest of all seven trajectories, totalling 3.85 million from 2000 to 2018 - equating to a 23.75% reduction (Table 4). Geographically, areas within this trajectiory are located exclusively in eastern and southern Europe - with the exception of areas in the former East Germany (Figure 4). Particularly concentrated in Balkan and Baltic countries and over-represented in Albania, Bulgaria, Latvia and Lithuania. Considering the persistent and rapid nature of decline, areas in this trajectory are most likely already experiencing the consequences of population decline.

*Accelerating Decline (6.61%)*: Areas which have undergone this trajectory display a tendency of rising annual rate of population decline, denoted by the transition from classes of moderate decline to decline. They have declined by a total of 1.57 million since the year 2000, representing an overall decline of 14.96%. Predominantly located in Southern Europe, chiefly in the Balkan countries of Croatia and Romania. Instances of *Accelerating Decline* can also be found in western Europe, in non-coastal Portugal and Spain.

*Diminishing Decline (8.19%)*: This trajectory describes a pattern of decelerating population decline, with a transition from our decline to moderate decline class, representing a decrease in rates of population decline. In total, areas experiencing this trajectory recorded a combined population loss of 7.82 million, or 15.04% (Table 4). These areas are largely located within eastern Europe in former member country members of the Soviet Union - Belarus, Estonia, Russia and Ukraine - and parts of east Germany.

*Persistent Moderate Decline (25.86%)*: This trajectory is defined by sustained moderate decline, with an annual rate of population change ranging between -0.3% to -0.99%, see Table 1. Very little movement between classes of population change is observed for this trajectory. This is the most common trajectory of population decline across Europe with a total of 180 sub-national areas, or 25.86%, experiencing this pathway of depopulation. Combined population losses in these areas since 2000 have totalled 7.86 million equating to a reduction of 9.3%. This trajectory is distributed across Europe in a total of 25 European territories and the predominant pathway of decline Austria, Finland, Hungary, Moldova, Poland, Romania, and Serbia.

*Accelerating Moderate Decline (15.37%)*: Areas in this trajectory first tended to experience a trajectory of accelerating decline comprising a transition from an extended period of stability to moderate population decline. These areas comprise a cluster of expansion of population decline in Europe in recent years. Population loss in areas is moderate, totalling just 1.95 million or 5.84% (Table 4). Distributed across the continent in a total of 17 territories, this trajectory is over-represented in Southern Europe and is the most common trajectory of decline in the countries of France, Greece, Italy, Romania, Spain, Slovenia, and Portugal. These areas are set to experience further decreases in the rate of population growth and thus accelerating the process of depopulation, similar to the process experienced by *Accelerating Decline* areas.

*Diminished Decline (15.23%)*: This trajectory describes a transitional pattern from the Moderate Decline to Stable classes. Here the annual rate of population decline is reduced to the point where areas are no longer considered in decline but rather in stability. This trajectory therefore represents the end of population decline, though a negative rate of annual population change is captured within the Stable state boundary (see Table 1). Areas in this trajectory have declined by a total of 2.81 million people, equating to a population reduction of 3.68% between 2000 and 2018. Geographically, these areas are predominantly found within central and eastern Europe and are most prevalent in Czechia, Slovakia, Sweden and the UK (Figure 4).

*Temporary Decline (18.82%)*: This trajectory is characterised by a sequential trend of Stable to Moderate Decline, followed by a short spell of population Growth, returning to within the bounds of the stability. Decline here is a temporary phenomenon, though severe enough to reduce the combined populations by 0.97 million or 4.03% from 2000 (table 4). Areas experiencing this trajectory are heavily concentrated in Germany. Of the 131 areas, 101 are within Germany. Such abrupt reversal of population decline could be linked to the influx of migrants during the Syrian refugee crisis (see Newsham & Rowe 2019).

Taken together, the identified trajectories reveal distinctive patterns of accelerating and stable rates of population decline, as well as trends of population reversal and temporary decline.

**Geographic Distribution of Decline Trajectories**

Analysing the geographic distribution of our trajectories of population decline across the continent reveals marked differences between the West-South and Central and North-East of Europe. *Accelerating Moderate Decline* and *Temporary Decline* trajectories are concentrated in the West-South and Central, particularly in Portugal, Spain, France, Germany, Italy and Slovenia. These patterns indicate that depopulation is either a recent or temporary phenomenon in these countries. In the North-East of Europe *Persistent Decline* and *Accelerating Decline* trajectories are dominant, reflecting a trend of fast paced population decline.

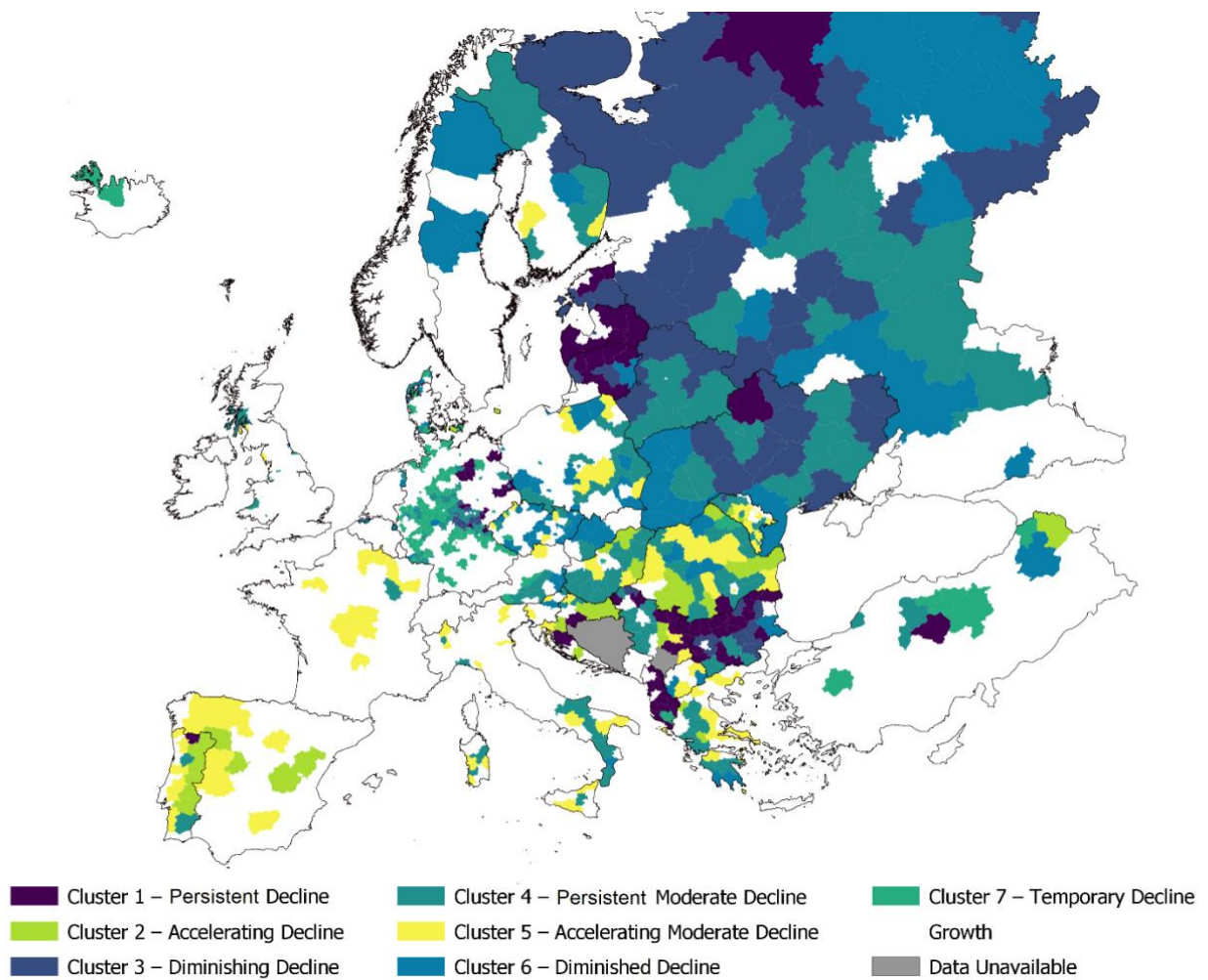

Figure 4 – Geographic Distribution of population decline trajectories

Additionally, examining the distribution of trajectories by country reveals differences in the prevalence of the pace and evolution of sub-national population decline (Figure 5). These differences reveal the co-existence of sub-national patterns of population decline in individual countries. Baltic nations are dominated by the *Persistent Decline* trajectory, depicting a picture of consistent fast paced population decline. Particularly, 80% of Latvian areas and 70% of Lithuanian areas are characterised by this decline trajectory. This is less evident in Estonia, where the trajectory *Diminishing Decline* is also well represented (50% of declining areas), indicating a slight divergence from the Baltic model of population change characterised by a deceleration of depopulation in parts of the nation. The dominance of the *Persistent Decline* trajectory is echoed in Balkan nations of Albania, Bulgaria, and to a lesser degree, Serbia, with 90%, 61% and 36% of areas experiencing this trajectory of decline, respectively. The *Diminishing Decline* trajectory is the dominant decline pathway in Belarus (50%), Ukraine (37.5%) and Russia (35.5%), suggesting that these nations transitioned through a period of rapid depopulation but this has started to decelerate from 2000. Elsewhere, in east and central Europe, *Persistent Moderate Decline* is widespread and is the prevailing trajectory of decline in Hungary (61.1%), Moldova (50%), Austria (45.5%), Poland (42.4%) and Romania (32.5%). Trajectories of accelerating population decline dominate the demographic landscape of Southern Europe. *Accelerating Moderate Decline* is the predominant pathway of decline in Spain (69.2%), Italy (53.3%), Slovenia (50%), Portugal (43.8%), and Greece (42.9%). Elsewhere in the South, within Croatia and Andorra, population decline has continued to accelerate with a rate of sub-national decline greater than in other nations in southern Europe. Here 100% and 53.3% of declining areas have experienced this trajectory of decline, respectively. Generally, sub-national areas in western and northern Europe show a propensity for decelerating population declines. This is demonstrated by a dominance of the *Diminished Decline* trajectory in Sweden (100%), the Netherlands (60%) and the UK (50%), as well as *Temporary Decline* in Iceland (100%), Germany (61.8%) and Denmark (39.1%).

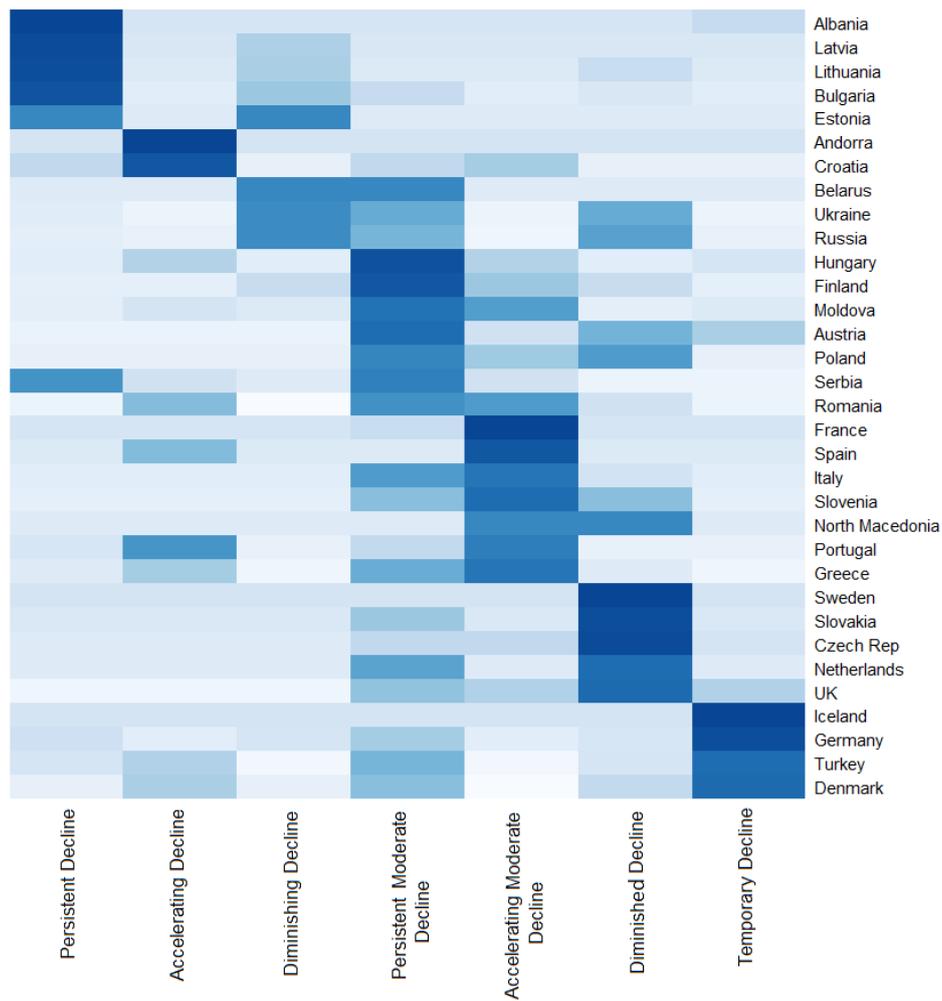

Figure 5 – Distribution of cluster trajectories in each European country

**Urban and Rural Population Decline**

While existing research suggests that countries follow a particular distribution of population decline spreading up across the urban hierarchy, our trajectories suggest that distinctive sets of patterns tend to coexist. To better understand differences in the temporal evolution of population decline across the rural-urban continuum, we classified subnational areas by population size (X-large, Large Mid-sized and Small) and rural population representation (Urban, Intermediate and Rural) as explained in the Method Section. Table 5 reports the number trajectories according to this classification.

|  | Urban | | | | | Intermediate | | | | | Rural | | | | |
| --- | --- | --- | --- | --- | --- | --- | --- | --- | --- | --- | --- | --- | --- | --- | --- |
| Trajectory Cluster | X-Large | Large | Mid-Sized | Small | Total | X-Large | Large | Mid-Sized | Small | Total | X-Large | Large | Mid-Sized | Small | Total |
| Stable Decline | - | 1 | - | 1 | 2 (2.27%) | 1 | 1 | 4 | 30 | 36 (11.73%) | - | - | 3 | 28 | 31 (10.30%) |
| Accelerating Decline | - | 1 | 1 | 1 | 3 (3.41%) | - | 1 | 5 | 5 | 11 (3.58%) | - | - | 5 | 27 | 32 (10.63%) |
| Diminishing Decline | 3 | 3 | 1 | - | 7 (7.95%) | 19 | 8 | 3 | 11 | 41 (13.36%) | - | - | 1 | 8 | 9 (2.99%) |
| Stable Mod Decline | 3 | 6 | 7 | 5 | 21 (23.86%) | 17 | 15 | 10 | 26 | 68 (22.15%) | 1 | 3 | 21 | 66 | 91 (30.23%) |
| Accelerating Mod Decline | 1 | 2 | 2 | 3 | 8 (9.09%) | 1 | 7 | 16 | 15 | 39 (12.70%) | - | 6 | 15 | 39 | 60 (19.93%) |
| Diminishing Mod Decline | 5 | 6 | 4 | 9 | 24 (27.27%) | 13 | 11 | 7 | 25 | 56 (18.24%) | 3 | 4 | 4 | 15 | 26 (8.64%) |
| Temporary Decline | 1 | 2 | 8 | 12 | 23 (26.14%) | - | 4 | 12 | 40 | 56 (18.24%) | - | - | - | 52 | 52 (17.28%) |
| In Decline | 13 | 21 | 23 | 31 | 88 | 51 | 47 | 57 | 152 | 307 | 4 | 13 | 49 | 235 | 301 |
| All Areas | 86 | 78 | 137 | 217 | 518 | 86 | 130 | 184 | 445 | 845 | 6 | 38 | 111 | 518 | 673 |
| In Decline (%) | 16.46 | 26.92 | 16.79 | 14.29 | 16.99 | 60.00 | 36.15 | 30.98 | 34.16 | 36.33 | 66.67 | 34.21 | 44.14 | 45.37 | 44.73 |

Table 5 – Distribution of cluster trajectories by area classification and size

We observe distinct differences in the occurrence and temporal pattern of population decline across rural, intermediate, and urban areas, and between areas of different population sizes. As expected, rural areas are more likely to experience a trajectory of population decline. However, Table 5 also reveals the unequal occurrence of depopulation across rural areas. Small and mid-sized rural areas are more likely to experience population decline than larger rural locales with 45.37%, 44.13%, and 34.21% of areas experiencing population decline respectively. Additionally, Table 5 shows the distinctive pathways of population decline undergone by rural areas. Not all rural areas have undergone a continuing pattern of depopulation. Instead, they also experience accelerating or decelerating trends of decline. Interestingly, we observe that small rural areas are typically oriented towards the acceleration of population decline, whereas larger rural locales demonstrate a greater propensity for deceleration. The depopulation trajectories experienced by rural areas also tend to differ from those observed in urban and intermediate locations. Rural areas are more likely to experience patterns of consistent and accelerating population declines. This evidence indicates that not all rural areas are the key driver of European population decline but predominantly those small in size.

Urban areas, on the other hand, are significantly less likely to experience population decline than rural and intermediate areas. Despite this they are represented in all trajectories. Generally, urban areas are more likely to have experienced decelerating or temporary population declines than persistent or accelerating pathways and depopulation in urban areas also seem to associate with population size. Unlike patterns of rural population decline, large urban areas seem more likely to undergo population decline than mid-sized and small urban locations, with 26.92%, 16.79%, and 14.29% of these areas experiencing decline, respectively.

The likelihood of intermediate areas experiencing population decline rather aptly lies in-between that of rural and urban areas. Similar to these other areas, Table 5 shows that the propensity for intermediate locales of different sizes to experience divergent pathways of depopulation. Particularly, we find that intermediate areas with small population sizes are more likely to experience accelerated population decline. Differently, larger intermediate areas show a tendency to experience trajectories of deceleration. On the whole, intermediate areas demonstrate the highest propensity for *Persistent Decline* and *Diminishing Decline*. Intermediate areas are thus seen as having a similar propensity for rapid population declines as rural areas, but also are as oriented towards deceleration as urban areas.

**Conclusion**

Population decline is set to overtake population growth and become the main trend of population change in most countries across Europe, with wide-ranging societal and economic implications. Yet, we know very little about the spatial and temporal dynamics of population decline across the urban-rural continuum. This study sought to address this gap and developed a unique methodological approach to analyse the trajectories of depopulation in a total of 696 sub-national areas across 43 countries in Europe over an 18-year period (2000-2018). Our findings revealed that depopulation has occurred across the rural-urban continuum; in rural, urban and intermediate areas of Europe. A key contribution of our work is the identification of a typology of European population decline distinguishing seven distinct pathways to depopulation. These pathways represent the systematic ways in which population decline has unfolded, temporally, since the year 2000. The pathways recognise persistent, temporary, accelerating and decelerating trajectories of depopulation and are distinguished by the extent, sequencing and timing of their transitions between various intensities of population decline. From our analysis we highlight three main findings.

Firstly, the most dominant pathway of population decline is *Persistent Moderate Decline*, though trajectories of accelerating and decelerating declines were well represented. Secondly, we identified the spatial concentration of the seven trajectories and noted patterns in individual countries and European sub-regions. Particularly, we observed persistent and rapid declines in the east, persistent but moderate declines in central Europe, accelerating declines in the south and decelerating population declines in the west. Population decline was demonstrated to be both a more widespread and longstanding feature of eastern Europe demography (Fihel & Okólski 2019, Coleman & Rowthorn 2011). Thirdly, we also revealed systematic differences in population decline across the rural-urban spectrum and between areas of different population size. We found that population declines in rural areas were oriented towards acceleration, signalling considerable challenges for these areas. Conversely in urban areas, the rate of population declines appears to be decelerating. We observed similar patterns between small and large populated areas, respectively, indicating that small and mid-sized rural areas are driving the process of population decline across Europe.

Our analysis also provided empirical evidence that can be used to enhance existing theories of population change across the urban-rural continuum. As proposed by the urban differentiation model, areas across this continuum are often assumed to follow a rigid progression through a set of predetermined stages (Geyer & Kontuly 1993). Yet, we showed that urban and rural areas of differing population size in individual countries do not transition through a single linear developmental pathway; that is, they do not follow a single pre-determined trajectory, in a similar fashion as suggested for the trends of fertility and mortality as anticipated by the demographic transition model. Local contingencies and past conditions act to create a set of distinct trajectories. We showed that a diverse number of depopulation trajectories can coexist, revealing simultaneous patterns of depopulation acceleration, stabilisation and reversal. We also showed that the geographical distribution of these trajectories follows particular patterns, which capture the differentiated impact of national and local economic, social and demographic forces.

Our typology of depopulation pathways has important policy implications. It can serve as a useful tool to identify at risk areas and areas of future concern in need of urgent policy intervention to mitigate or prevent the negative consequences associated with population decline. Our analysis revealed that population decline in the east of Europe have been more severe and sustained than elsewhere in the continent. At the same time, we identified areas that demonstrated the capacity to reverse trajectories of population decline in Germany, Sweden and the Netherlands, and now highlight the potential for further research into these areas to understand the processes underpinning this reversal in depopulation. We anticipate considerable value to such research in regards to developing policy measures that can be applied across the continent. Policy efforts to reverse population decline or mitigate the negative consequences of this trend on the economy and labour market should be concentrated particularly on these geographic areas.

We anticipate multiple avenues for further research in relation to our typology of European population decline. Particularly, future work should focus on expanding this analysis into the future as new data become available, to assess the potential increase in severity of depopulation across Europe and geographical spread throughout the rural-urban continuum. Additionally, Future research should also investigate the underlying the causes driving the observed spatio-temporal dynamics of depopulation. Understanding the relative importance of the ways fertility, mortality, internal and international migration contribute to shape local patterns of population decline would be of great importance to identify appropriate policy interventions.